\documentclass[12pt,preprint]{aastex}

\shorttitle{The Pattern Speeds of NGC 2523 and NGC 4245}
\shortauthors{Treuthardt, Buta, Salo, \& Laurikainen}

\begin{document}


\title{The Kinematically Measured Pattern Speeds of NGC 2523 and NGC 4245}


\author{P. Treuthardt\altaffilmark{1}, R. Buta\altaffilmark{1}, 
H. Salo\altaffilmark{2}, and E. Laurikainen\altaffilmark{2}}\altaffiltext{1}
{Department of Physics and Astronomy, University of Alabama, Box 870324, 
Tuscaloosa, AL 35487}\altaffiltext{2}{Division of Astronomy, Department of 
Physical Sciences, University of Oulu, Oulu, FIN-90014, Finland}


\begin{abstract}

We have applied the Tremaine-Weinberg continuity equation method to derive
the bar pattern speed in the SB(r)b galaxy NGC 2523 and the SB(r)0/a galaxy
NGC 4245 using the Calcium Triplet absorption lines. These galaxies
were selected because they have strong inner rings which can be used as
independent tracers of the pattern speed. The pattern speed of
NGC 2523 is 26.4 $\pm$ 6.1 km s$^{-1}$ kpc$^{-1}$, assuming an
inclination of 49.7$^{\circ}$ and a distance of 51.0 Mpc. The pattern speed of 
NGC 4245 is 75.5 $\pm$ 31.3 km s$^{-1}$ kpc$^{-1}$,
assuming an inclination of 35.4$^{\circ}$ and a distance of 12.6 Mpc. The
ratio of the corotation radius
to the bar radius of NGC 2523 and NGC 4245 is 1.4 $\pm$ 0.3 and
1.1 $\pm$ 0.5, respectively.
These values place the bright inner rings near and slightly inside the
corotation radius, as predicted by barred galaxy theory. Within the
uncertainties, both galaxies are found to have fast bars that likely indicate 
dark halos of low
central concentration. The photometric properties, bar strengths, 
and disk stabilities of both galaxies are also discussed.

\end{abstract}




\section{Introduction}

The presence of barred structures in disk galaxies is fairly common.
The fraction of spiral galaxies containing a strong bar
is 30$\%$ in the optical (de Vaucouleurs 1963) and is nearly double
that when counted in the near-infrared
(Eskridge et al. 2000; Hernandez et al. 2005). The bar pattern speed
($\Omega_{P}$), or the rate at which the bar rotates, is one of the
main factors influencing the morphology and dynamical structure of
the host galaxy. Although
this parameter is important in the secular evolution of galaxies
(e.g., Lynden-Bell \& Kalnajs 1972; Kalnajs 1991; Kormendy \& Kennicutt
2004), it has only been determined for
a small number of galaxies. The common methods of determining pattern speed
involve either identifying morphological features, such as rings, with
resonance radii
(Buta \& Combes 1996), looking for residual patterns in the
velocity field (Canzian 1993; Purcell 1998), looking for phase crossings
of $B$
and $I$-band spirals (Puerari \& Dottori 1997) or H$\alpha$ and CO
arms (Egusa et al. 2006), or recreating the observed morphology through
dynamical models
(e.g., Salo et al. 1999; Rautiainen, Salo, \& Laurikainen 2005).
A new method based on the theoretically-predicted azimuthal phase
shift between the potential
and density in a galaxy with a global, self-consistent mode has
been proposed by Zhang \& Buta (2007), who also summarize other
methods to locate corotation and estimate pattern speeds.

Most of these methods give only indirect estimates of the bar
pattern speed, either through model assumptions or because they
locate resonances directly and require a rotation curve to estimate
a pattern speed. A more direct approach is through the kinematic method
derived by Tremaine \& Weinberg (1984, hereafter TW). TW determined
that the pattern speed of a bar can be estimated from the
luminosity weighted mean
line-of-sight velocities, $\langle$$V$$\rangle$, and luminosity weighted mean
positions, $\langle$$X$$\rangle$, of a tracer that obeys the
continutity equation. These quantities are to be measured along lines
parallel to a barred galaxy's major axis. The galaxy's inclination, $i$,
is also required to determine $\Omega_{P}$. The quantities are related as
follows,
\begin{equation}\Omega_{P} \sin{i} = \frac{\langle V \rangle}{\langle X \rangle}.\end{equation}

A requirement of the TW method is that the continuity equation be satisfied.
This led to SB0 galaxies being the first objects to which the method was
applied (e.g., Merrifield \& Kuijken 1995; Gerssen et al. 1999).
These galaxies can
have strong bar patterns but lack the dust and star formation that complicate
similar measurements for later-type systems. Since the continuity equation 
requires that the tracer
be something that is neither created nor destroyed, significant star formation
would violate the equation. This is the reason that applications to spirals
have been
more limited. Hernandez et al. (2005) discuss the application of the TW method
to atomic, molecular, and ionized gas phases in spiral galaxies.

The main goal of previous SB0 studies was to use the
TW method to measure the ratio of
the corotation radius (where the circular angular
velocity $\Omega$ = $\Omega_p$)
to the bar radius. 
If the ratio is between 1.0 and 1.4,
a bar is said to be "fast," 
while if greater than 1.4, a bar is said to be "slow". 
Debattista \& Sellwood (2000) argued that fast bars exist in halos with 
a low central concentration since the bar rotation rate would rapidly 
decrease due to dynamical friction with the halo. Certain galaxy 
models from Athanassoula (2003) also show this correlation between fast bars
and halos of low central concentration. 
A drawback of SB0 
galaxies is their simplicity. Apart from the ill-defined "ends" of the bar, 
there are no gaseous features that might be tied to the pattern speed that 
could be used to evaluate other implications of the method.


Ringed SB0/a galaxies offer SB0-like galaxies with conspicuous
rings of gas that can be tied to specific resonances through numerical
models. In addition, application of the TW method at longer wavelengths
than previous studies allows the possibility of measuring pattern
speeds even in intermediate-type spirals which are more affected by dust. 
If such spirals also have rings, then the resonance idea can be tested in 
them as well.

The goal of our study is to determine the pattern speed of two resonance ring
galaxies by applying the TW method in order to examine the central 
concentration of the dark matter halo as well as the possible resonance 
identifications of the rings. This paper details our TW analysis of the barred
galaxies NGC 2523 and NGC 4245, two excellent cases which show strong inner
rings. The observations are summarized in section 2. The subsequent analysis
of the data is described in section 3, while the measured pattern speeds and
their significance is discussed in section 4. Section 5 discusses
the potential testing of barred spiral theory in the future using the
findings from this paper.

\section{Observations}

The barred galaxies, NGC 2523 and NGC 4245,
were selected for this project because of
their strong inner ring features, accessibility, inclination, and
preferential orientation
of the bar axis to the galaxy major axis. 
Figure 1 shows $B$-band images obtained with the Nordic Optical Telescope
in 2003 January and 2004 January, respectively (Laurikainen et al. 2005). 
The images are in the
units of mag arcsec$^{-2}$ used in the de Vaucouleurs Atlas of Galaxies
(Buta, Corwin, \& Odewahn 2007, hereafter BCO). NGC 2523 is a member of a 
small nearby group called LGG 154, whose mean redshift is 3674 km s$^{-1}$
(Garcia 1993). The distance we adopt for NGC 2523 is 51.0 Mpc 
(Kamphuis et al. 1996) with a Hubble constant of 70 km s$^{-1}$ Mpc$^{-1}$. 
It has an absolute 
$B$-band magnitude of $-$21.6 (derived from BCO) and a revised de Vaucouleurs 
type of 
SB(r)b. The inner ring of this galaxy is a conspicuous closed feature, a rare 
true ring rather than a pseudoring. Outside the inner ring, the spiral pattern 
is multi-armed and no outer pseudoring is formed. NGC 4245 is a low luminosity 
member of the Coma I galaxy group, known to harbor a significant number of HI
deficient galaxies (Garc\'{i}a-Barreto et al. 1994). The group (LGG 279) 
includes 17 galaxies listed by Garc\'{i}a (1993) and has a mean redshift of
974 km s$^{-1}$. The distance we adopt for NGC 4245 is 12.6 Mpc 
(Garc\'{i}a-Barreto et al. 1994) with an 
H$_{0}$ value of 70 km s$^{-1}$ Mpc$^{-1}$. NGC 4245 has an absolute $B$-band 
magnitude of $-$18.0 (derived from BCO) and a revised de Vaucouleurs type of 
SB(r)0/a. In addition to a gaseous inner ring made
of tightly wrapped spiral arms, NGC 4245 has a very regular nuclear ring,
but lacks an outer ring that is often present
in such a galaxy. Garc\'{i}a-Barreto et al. (1994) found that NGC 4245 is one
of the more
HI deficient members of the Coma I group, suggesting that its lack of an
outer ring is due to gas stripping.

In each galaxy, the bar is well-defined
and the inner ring deprojects into an intrinsic elliptical shape aligned
parallel to the bar (BCO). The deprojected ring major axis radii and axis
ratios are
35\rlap{.}${\arcsec}$3 (8.7 kpc) and 0.74 for the inner ring of NGC 2523,
40\rlap{.}${\arcsec}$6 (2.5 kpc) and 0.77 for the inner ring of NGC 4245, and
4\rlap{.}${\arcsec}$8 (0.29 kpc) and 0.92 for the nuclear ring of NGC 4245.
Color index maps (also in BCO) indicate that the rings are
narrow zones of star formation. The inner ring of NGC 2523 has a somewhat
asymmetric
star formation distribution, while the inner and nuclear rings of NGC 4245
are both well-defined symmetric regions of enhanced blue colors. In the
near-infrared, the star forming rings of NGC 4245 are much
less conspicuous and the galaxy very much resembles an SB0 galaxy.
These observed characteristics of the galaxies' rings are similar to
those of gaseous resonance
rings formed in test-particle simulations of barred galaxies (see review
by Buta \& Combes 1996). For this reason, NGC 2523 and NGC 4245 are ideal
for our test.

We obtained long-slit spectra of NGC 2523 and NGC 4245 on the nights of
2006 January 18 and 19 using the RC Spectrograph on the KPNO Mayall 4 meter
telescope.
The T2KB CCD was used as a detector and was configured with the KPC-18C
grating
and a 2$\arcsec$ x 5.4$\arcmin$ slit which provided a spectral resolution of
2.44$\AA$. The region encompassing the Calcium Triplet (8498, 8542, 8662$\AA$
rest wavelengths; Dressler 1984) was observed. The benefits of observing in
this region are
the reduced influence of extinction and the strength of the absorption lines.
The main disadvantage is that there are a significant number of night sky
emission
lines in the same wavelength domain. The long-slit positions were set both
coincident with and offset parallel to the major axis of each galaxy
(see Figure 2). The total exposure time for each offset slit position
was 9000 seconds. The exposure times for positions along the major axis of
NGC 2523 and NGC 4245 were 6000 and 7500 seconds, respectively (see Table 1).
Spectra were also taken of two K-giant stars, HD 106278 and HD 109281, which
served
as velocity standards. The spectra were reduced, combined, and sky-subtracted
using standard
IRAF\footnote{IRAF is distributed by the National Optical Astronomy
Observatories,
which are operated by AURA, Inc., under cooperative agreement with the
National
Science Foundation.} routines. The sky levels were determined from rows
in the outer parts of the slit in each case. Imperfect subtraction of the 
night sky lines is a significant source of uncertainty mainly for NGC 2523,
due to its unfavorable redshift. It should be noted also that the atmospheric
observing conditions
were less than ideal during our two observing nights due to intermittent 
cloud cover and high wind speeds that required the dome to be closed on one 
occasion. As a result, more observing time was spent to reduce noise on the 
outer spectral positions than the central positions. The outer positions 
correspond to the endpoints of a line fitted to data 
plotted in the $\langle V \rangle$-$\langle X \rangle$ plane. Accuracy in the 
endpoints provides a more accurate slope determination of a line fit to the 
points.

In addition to the spectra, we have used 2.15$\mu$m $K_s$-band images,
as well as standard $B$ and $V$-band images,
obtained with the 2.5-m Nordic Optical Telescope (NOT) for some of our 
analysis. Details of these observations are provided by Laurikainen et al.
(2005) and Buta et al. (2006).

\section{Analysis}

\subsection{Bar, Bulge, and Disk Properties}

Detailed isophotal analysis of the deep NOT $B$ and $V$-band images
was used to derive photometric orientation parameters of the two
galaxies listed in Table 2. The $K_s$-band image is not as deep as
the optical images in each case, and thus the optical parameters
were used to deproject the $K_s$-band images. Deprojection was
facilitated by two-dimensional decomposition using the approach
outlined by Laurikainen et al. (2004, 2005). This is a multicomponent
code, which for our purposes was used with a
Sersic $r^{1\over n}$ function to parametrize the bulge, an
exponential to parametrize the disk, and a Ferrers function to
parametrize the bar. For the bulge, $n$=4 corresponds to a
de Vaucouleurs $r^{1\over 4}$ law while $n$=1 corresponds to
an exponential. The decomposition parameters listed in Table 2
for NGC 4245 are from Laurikainen et al. (2005). In both galaxies,
the Sersic index $n$ is close to 1.0, implying bulges that more
exponential-like than the de Vaucouleurs profile.

Bar strengths were estimated using the gravitational torque
approach (Laurikainen \& Salo 2002), where $Q_b$ denotes the maximum of the
tangential force 
normalized to the mean axisymmetric radial force. This approach uses a 
deprojected $K_s$-band image that is converted to a gravitational 
potential assuming a constant mass-to-light ratio. The deprojections were 
performed by subtracting
the bulge model from the total image, deprojecting the residual bar and
disk components, and then adding the bulge back as a spherical component.
The spiral arm torques are small for NGC 4245, so a bar-spiral separation
analysis was not needed to determine the bar strength $Q_b$. In this case, 
$Q_b$ can be taken as the total nonaxisymmetry strength in the galaxy given 
by Laurikainen et al. (2005).
For NGC 2523, it was necessary to do a bar-spiral
separation analysis in order to remove the effects of spiral arm
torques. The procedure we used for this is outlined
by Buta, Block, \& Knapen (2003) and Buta et al. (2005).
We found that a single Gaussian represents the radial relative
Fourier intensity profiles of the bar of NGC 2523 fairly effectively,
allowing a clean separation of the bar from the spiral.
For NGC 4245, Buta et al. (2006) show that the relative
Fourier intensity profiles of the bar require a double gaussian fit,
owing to the coexistence of the primary bar with an aligned
oval.

The resulting bar strengths listed in Table 2 are 0.55 for
NGC 2523 and 0.18 for NGC 4245, assuming the vertical density
profile is exponential. For NGC 2523, we assumed a vertical
scale height of $h_z=h_r/5$, while for NGC 4245, we assumed
$h_z=h_r/4$, based on the empirical correlation derived by de Grijs (1998) 
between $h_r/h_z$ and morphological type.
The bar in NGC 2523 is clearly exceptional in strength, while
that in NGC 4245 barely merits the SB classification.

\subsection{Mean Positions and Line-of-Sight Velocities Along the Slit}

The luminosity weighted mean position along each slit position was determined
by summing the two-dimensional data in the spectral direction. The resulting
luminosity profiles versus slit position $x$ are shown in the upper panels of 
Figures 3 and 4. The mean positions were calculated
from these profiles and are marked on the figures (vertical dashed lines)
as well as recorded in
Table 1. The errors in $\langle$$X$$\rangle$ are negligible compared with
those in $\langle$$V$$\rangle$.

The lower panels of Figures 3 and 4 show the line of sight velocity profiles
versus the slit position $x$. The radial velocities for each slit position were
determined by cross correlating the galaxy spectra with the spectra of the
standard
stars using the IRAF routine XCOR (e.g. Tonry \& Davis 1979) and applying 
a heliocentric correction. The XCOR
cross correlation routine assumes
a galaxy spectrum is a convolution of a stellar spectrum
with a Gaussian which describes the line-of-sight velocity dispersion of the
galaxy's stars.
A stellar template spectrum is cross correlated with the galaxy spectrum to
produce a function with a
peak at the redshift of the galaxy and with a width corresponding to the
dispersion of the galaxy.
Peaks in the cross correlation function are identified and fitted by
parabolas to obtain their position
and width. XCOR was used instead of FQUOT, a Fourier quotient routine,
because it was more stable for our purposes.
Due to the strength and number of night sky emission
lines in the galaxy spectra, each absorption line in the Calcium Triplet was
individually
cross correlated to the corresponding line found in the standard star
spectra. This was possible because the Calcium Triplet lines are
well-separated and can be treated individually.

The horizontal dashed lines in Figures 3 and 4 show the luminosity-weighted
mean velocities for each slit position for NGC 4245 and three of the slit
positions for NGC 2523. The results for NGC 2523 were especially difficult 
to extract because of the stronger night sky emission line contaminations in 
that case, due to the higher redshift as noted. The luminosity-weighted
mean velocities of each slit position for both galaxies are also listed in 
Table 1. The systemic velocities we find for both galaxies agree with 
published estimates within the errors (i.e. Catinella et al. 2005; 
Springob et al. 2005).

In Figures 5 and 6, we show examples of line profiles derived from co-adding 
the 8542$\AA$ rest wavelength spectra for each offset position of each galaxy. 
The line profiles were derived by fitting four Gaussian components to the 
observed profile, similar to what has been done in previous TW studies 
(e.g. Merrifield \& Kuijken 1995; Gerssen et al. 1999). The 
luminosity-weighted 
mean velocities derived from these example profiles (see captions in Figures 5 
and 6) support the same sense of rotation as those derived from XCOR, though 
they are found to be larger than the XCOR derived velocities (Table 1). 
In a TW analysis of four galaxies, Gerssen et al. (2003) concluded that the 
obtained pattern speeds did not differ significantly when either 
technique was applied to determine $\langle$$V$$\rangle$. An advantage of using
XCOR is that velocities and velocity dispersions are 
determined along each slit, allowing the derivation of rotation curves and the 
evaluation of disk stability properties.

\subsection{Circular Velocities, Velocity Dispersions, and Disk Stability}

Rotation curves of the two galaxies can be easily derived from the velocities
extracted by XCOR of the different slit positions. The deprojected circular
velocities were determined
with the assumption that NGC 2523 has an inclination of 49.7$^{\circ}$ 
and NGC 4245 has an 
inclination of 35.4$^{\circ}$ [derived from 
$\langle$$q$$\rangle$ $(disk)$ in Table 2 and assuming an intrinsic oblate 
spheroid axis ratio ($c/a$) of 0.2 (Schommer et al. 1993)].
These rotation curves are shown in Figure 7 along with published estimates 
of the maximum gaseous rotation velocities in each case (horizontal lines). 
We notice 
that the inner 20$\arcsec$ of our rotation curve for NGC 2523 resembles that 
of H\'{e}raudeau et al. (1999), which was determined through Mg $b$ 
absorption spectroscopy. The velocity
dispersions outputted by XCOR were corrected for an instrumental dispersion of
30 km s$^{-1}$ and are shown in Figure 7 as well. We have assumed that
$\sigma_{r}$=$\sigma_{\phi}$=$\sigma_{z}$ in which case the observed velocity 
dispersion 
is equal to $\sigma_{r}$. The velocity dispersions we measure in 
NGC 2523 are approximately a third of those found by 
H\'{e}raudeau et al. (1999) in the same radial range. We find that the 
velocity dispersions in NGC 4245 appear constant across the observed portions 
of the galaxy. This behaviour is not unprecedented in SB0 galaxies.
Constant velocity dispersion values were also found across the SB0 galaxy
NGC 4596 (Bettoni \& Galetta 1997). If we assume that 
$\sigma_{r}=\sigma_{\phi}$, $\sigma_{z}$=0, and the observed velocity
dispersion is $\sigma_{r}$ $\sin{i}$, then the velocity dispersions plotted in 
Figure 7 would be increased by a factor of 1/$\sin{i}$ for each galaxy. 
For the 
inclinations we assume, this corresponds to an increase in the 
velocity dispersions by a factor of 1.31 for NGC 2523 and 1.73 for NGC 4245.

The stability of the galaxy disks can be evaluated
with knowlege of the velocity dispersions and a few estimated parameters.
Toomre (1964) showed that a two-dimensional galaxy disk is stable  
against axisymmetric perturbations when
\begin{equation}Q \equiv \frac{\sigma_{r}\kappa}{3.36G\Sigma} > 1,\end{equation}
where $\sigma_{r}$ is the radial velocity dispersion, $\kappa$ is the
epicyclic frequency,
$G$ is the gravitational constant, and $\Sigma$ is the surface mass density
of the disk.

Gaseous rotation curves are needed to determine $\kappa$ since they do not 
suffer from the same effects of velocity dispersion support as stellar 
rotation curves. Because gaseous rotation curves of these galaxies are not 
available, we assume that the rotation velocity of the gas component is 
constant and equal to the maximum gaseous rotational velocity ($V_{max}$) 
obtained in the literature from observed HI line widths 
(i.e. Kamphuis et al. 1996; Garc\'{i}a-Barreto et al. 1994). In this case, 
$\kappa$ is equal to $\sqrt{2} V_{max}/r$.

In order to derive Toomre $Q$ for the stellar component alone, we have
used surface brightness profiles in conjunction with a color-dependent
mass-to-light ratio formula from Bell \& de Jong (2001) to estimate
$\Sigma$.
We used the NOT $V$ and $K_S$-band images to obtain azimuthally averaged 
surface brightness profiles of the galaxies (Figure 8). Published 
photoelectric aperture photometry from
Longo \& de Vaucouleurs (1983) was used to calibrate the $V$-band
images, while 2MASS photometry within a 14$\arcsec$ aperture
from the NASA/IPAC Extragalactic Database (NED) was used to calibrate
the $K_S$-band images.
Mass-to-light ratios were derived from the $V-K_S$ color profiles
after correction for Galactic extinction values from NED and
using $log(M/L)_K=-1.087+0.314(V-K_S)$ from Table 1 of Bell \& de Jong
(2001). (This ignores the generally small difference between $K$ and $K_s$.)
Surface mass density profiles were derived by converting
the azimuthally-averaged surface brightness profiles $\mu_{K_S}$ into
solar $K$ luminosities per square parsec using an absolute magnitude
$M_K(\odot)$=3.33 from Worthey (1994), and multiplying the values by $(M/L)_K$.

Figure 9 shows our determinations of the lower limit values of $Q$ versus 
radius for both galaxies (as seen in Figure 6 of Kormendy 1984). In this 
figure 
we assume that $\sigma_{r}$=$\sigma_{\phi}$=$\sigma_{z}$ and the observed 
velocity dispersion is equal to $\sigma_{r}$. For NGC 2523, $Q$ ranges from 
0.7 $\pm$ 0.1 to 1.6 $\pm$ 0.6. $Q$ falls below 1 from approximately 
4$\arcsec$ to 6$\arcsec$ and 32$\arcsec$ to 34$\arcsec$.
For NGC 4245, $Q$ ranges from 0.6 $\pm$ 0.1 to 5.0 $\pm$ 0.7 and falls below 1 
from approximately 2$\arcsec$ to 4$\arcsec$. 
The upper limit values of $Q$ are found from our 
assumption that $\sigma_{z}$=0. In this case, $Q$ ranges from 1.0 $\pm$ 0.2 to 
2.1 $\pm$ 0.8 for NGC 2523 and reaches 1 at approximately 6$\arcsec$. 
For NGC 4245, $Q$ ranges from 1.1 $\pm$ 0.2 to 8.6 $\pm$ 1.1. 
The results imply marginal stability for 
NGC 2523 and much higher stability for NGC 4245. This could explain the 
latter's much smoother light distribution.

\section{Results}
\subsection{Pattern Speeds}

From the data in Table 1, it is straightforward to calculate the bar pattern
speeds of the
two sample galaxies. Figure 10 shows the plots of $\langle$$V$$\rangle$ versus
$\langle$$X$$\rangle$, where the slope of the line fitting the data points is
$\Omega_{P}$ $\sin{i}$. The slope and 
corresponding 1$\sigma$ error that best fits the NGC 2523 data is
$\Omega_{P}$ $\sin{i}$ = 5.0 $\pm$ 1.2 km s$^{-1}$ arcsec$^{-1}$. If we
consider that the
galaxy is inclined 49.7$^{\circ}$ and is at a distance of 51.0 Mpc,
then $\Omega_{P}$ = 26.4 $\pm$ 6.1 km s$^{-1}$ kpc$^{-1}$.
The slope and corresponding 1$\sigma$ error that best fits the NGC 4245 data
is $\Omega_{P}$ $\sin{i}$ = 2.7 $\pm$ 1.1 km s$^{-1}$ arcsec$^{-1}$. If we
consider that this
galaxy has an inclination of 35.4$^{\circ}$ and is at a distance of
12.6 Mpc, then $\Omega_{P}$ = 75.5 $\pm$ 31.3 km s$^{-1}$ kpc$^{-1}$.

The $\chi^{2}$ value for the linear fit of the NGC 2523 data is 2.2 while for
NGC 4245 it is 0.1, each with 3 degrees of freedom
(Merrifield \& Kuijken 1995). The $\chi^{2}$
values imply that the observations from different slit positions are all
consistent with the single bar pattern speed for each galaxy as described
above. This also implies that the error
analysis returns a realistic measure of the uncertainty in each of the
$\langle$V$\rangle$
estimates. It is noteworthy that NGC 4245 is the first SB0 galaxy with strong
resonance rings to have its pattern speed measured by the TW method.

\subsection{Maximum Disks and Frequency Curves}

A maximum disk, interpreted to be correlated with a fast bar, can be
determined from the distance independent ratio
$\cal R$ $\equiv$ $R_{CR}$/$R_{B}$, where $R_{CR}$ is the corotation radius
and $R_{B}$ is the bar semimajor
axis radius. Fast bars occur in the 1.0 $\leq$ $\cal R$ $\leq$ 1.4 regime
(Debattista \& Sellwood 2000), while bars are considered
slow when $\cal R$ $>$ 1.4. Contopoulos (1980) has concluded that 
self-consistent bars cannot exist when $\cal R$ $<$ 1.0, though Zhang 
\& Buta (2007) have 
argued to the contrary. With the assumed flat rotation curve, $R_{CR}$ is 
found by simply dividing
the maximum circular velocity ($V_{max}$) by $\Omega_{P}$. In the case of NGC
2523, if we assume an
inclination of 49.7$^{\circ}$ then $V_{max}$ is
294 km s$^{-1}$ (Kamphuis et al. 1996) and the deprojected bar length of the
galaxy can be visually estimated as 33.5$\arcsec$. Taking the inclination of
NGC 4245 to be 35.4$^{\circ}$, $V_{max}$ is 199 km s$^{-1}$
(Garc\'{i}a-Barreto et al. 1994) and the deprojected bar
length can be estimated to be 38.1$\arcsec$ by visual inspection. $\cal R$
for NGC 2523 is then
1.4 $\pm$ 0.3 and for NGC 4245 is 1.1 $\pm$ 0.5. The errors given should be
considered as minimum errors of $\cal R$. The $V_{max}$ values are shown as
compared to our derived rotation velocities in Figure 7.

With knowledge of the maximum circular velocity of a galaxy, it is
straightforward
to derive the familiar Lindblad precession frequency curves that show how
resonance locations vary with
angular velocity in the linear (epicyclic) approximation.
Overplotting the pattern speed of the galaxy allows one to
predict the
radius at which the resonances and possible resonance features occur.
This is shown in Figure 11. One caveat is that these curves are not
reliable at small radii because we have assumed only a single
rotation velocity. In reality, the rotation curves would rise more slowly
to a maximum, such that $\Omega-\frac{\kappa}{2}$ would show a finite
maximum. We could evaluate
the resonance identification of the nuclear ring of NGC 4245 only with
improved rotation information
in the central few kpc. Though the precision of
our $\Omega_p$ estimates is not very high, we can say that the inner rings
of our two galaxies lie
close to and inside the bar corotation radius. This is consistent with the
barred spiral theoretical studies of Schwarz (1984) and Rautiainen \& Salo
(2000). The curves for NGC 4245 do suggest that
the outer Lindblad resonance (OLR) should lie within the visible disk. Within
the uncertainties, the outer spiral pattern of NGC 2523 extends to the OLR.

\section{Future Work}

It remains to be seen how the kinematic bar pattern speeds measured through
the use
of the TW method compare to the dynamical bar pattern speeds derived through
simulations (e.g., Rautiainen, Salo, \& Laurikainen 2005).
If the values derived by the two methods concur, it would be a
strong indicator that the theory describing bar patterns
agrees with observations. This is important because galaxy modeling can be
applied to a wide range of galaxies, while the TW method is limited to
galaxies
oriented at a modest inclination with intermediate bar-to-major-axis
position angles. NGC 4245 is an excellent candidate for such a test due to
the good statistical fit of the measured pattern speed as well as the
strong multiple resonance ring features (i.e. nuclear and inner rings). The
resonance features apparent in this galaxy will help to constrain the
dynamical models used to determine the bar pattern speed.

P. Treuthardt and R. Buta acknowledge the support of NSF Grant AST050-7140. 
H. Salo and E. Laurikainen acknowlege the Academy of Finland for support.
This research made use of the NASA/IPAC Extragalactic Database (NED), which
is operated by the Jet Propulsion Laboratory, California Institute of
Technology, under contract with NASA.

\clearpage


\clearpage

\begin{deluxetable}{lccccc}
\tabletypesize{\scriptsize}
\tablewidth{0pc}
\tablecaption{Galaxy Data\tablenotemark{a}}
\tablehead{
\colhead{Galaxy}
&\colhead{Offset}
&\colhead{T$_{exp}$}
&\colhead{$\langle$$X$$\rangle$} &\colhead{$\langle$$V$$\rangle$}\\
\colhead{1}
&\colhead{2}
&\colhead{3}
&\colhead{4} &\colhead{5}
}

\startdata
NGC 2523  & NW 17$\arcsec$ & 9 x 1000 & 138.1 & 3505.4 $\pm$ 18.9 \cr
         & NW 9$\arcsec$  & 9 x 1000 & 144.9 & 3498.4 $\pm$ 28.6 \cr
         &  0$\arcsec$    & 6 x 1000 & 150.5 & 3496.5 $\pm$ 75.8 \cr
         & SE 17$\arcsec$ & 9 x 1000 & 164.9 & 3636.0 $\pm$ 24.8 \cr
NGC 4245  & E 17$\arcsec$  & 9 x 1000 & 144.2 & 967.4 $\pm$ 14.5 \cr
         &  0$\arcsec$    & 7 x 1000 + 500 & 152.6 & 928.4 $\pm$ 52.9 \cr
         & W 17$\arcsec$  & 9 x 1000 & 163.0 & 917.1 $\pm$ 14.9  \cr \enddata
\tablenotetext{a}{Explanation of columns: (1) Galaxy name; (2) direction and amount the slit was offset parallel to the galaxy major axis; (3) amount of exposure time at this slit position in seconds; (4) luminosity weighted mean position along the slit in arcseconds; (5) luminosity weighted mean line-of-sight velocity in km s$^{-1}$.}
\end{deluxetable}

\clearpage

\begin{deluxetable}{lccc}
\tabletypesize{\scriptsize}
\tablewidth{0pc}
\tablecaption{Basic Galaxy Properties}
\tablehead{
\colhead{Parameter\tablenotemark{a}}
&\colhead{NGC 2523}
&\colhead{NGC 4245\tablenotemark{b}}
}

\startdata
$\langle$$q$$\rangle$ (disk)                & 0.665 $\pm$ 0.005               &  0.823 $\pm$ 0.011 \\
$\langle$$\phi$$\rangle$ (disk)             & 60.7 $\pm$ 1.1                  &  174.1 $\pm$ 2.2   \\
Sersic index $n$ (bulge)    & 1.09                          &  1.33            \\
$q$ (bulge)                 & 1.0                           &  1.0             \\
$h_r$ (disk)                & 32$\rlap{.}$$\arcsec$1   & 25$\rlap{.}$$\arcsec$9\\
$B/T$                       & 0.07                          & 0.20             \\
$Q_b$                       & 0.55                          & 0.18             \\
\enddata
\tablenotetext{a}{The listed photometric parameters are the axis ratio $q$, position angle $\phi$, 
radial scale length $h_r$, bulge fraction $B/T$, and bar strength $Q_b$.}
\tablenotetext{b}{Parameters are from Laurikainen et al. 2005.}
\end{deluxetable}

\clearpage

\begin{figure}
\figurenum{1}
\includegraphics[angle=0,height=125mm]{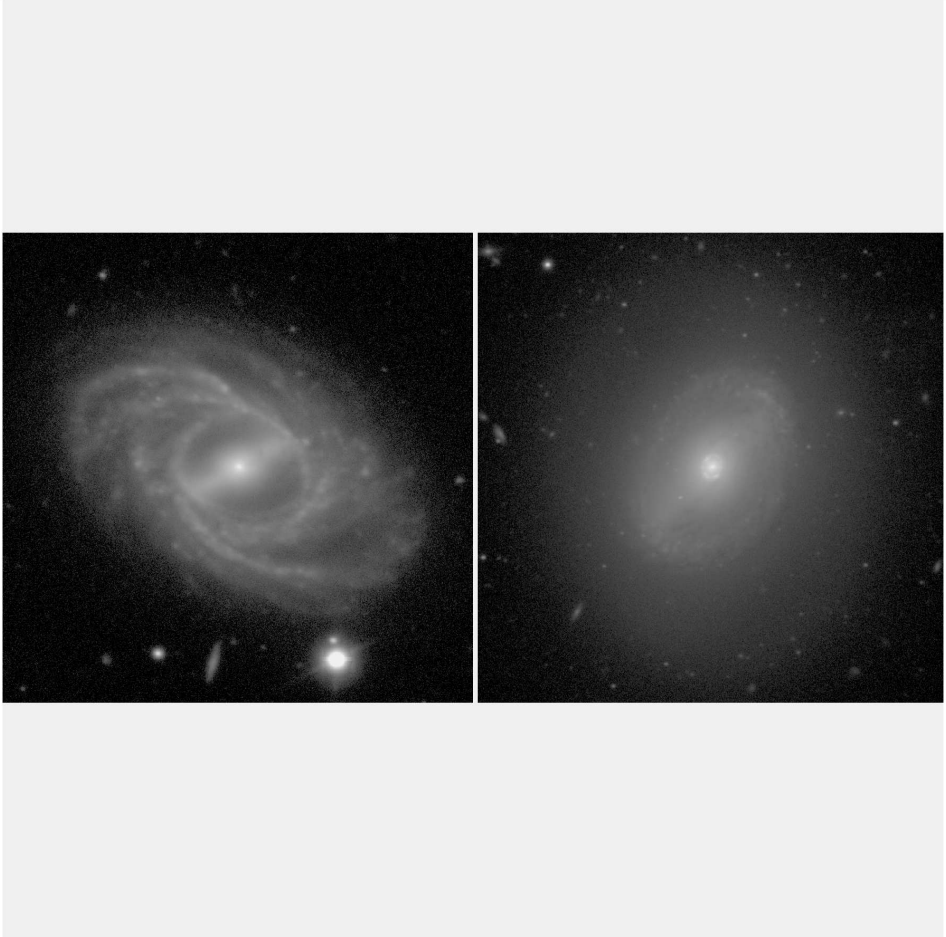}
\label{sample}
\caption{$B$-band images of NGC 2523 (left) and 4245 (right). 
These were obtained with the Nordic Optical Telescope in 2003
and 2004 (Laurikainen et al. 2005) and are presented as illustrated in the 
de Vaucouleurs
Atlas of Galaxies (Buta, Corwin, and Odewahn 2007). North is
at the top and east is to the left in each case.}
\end{figure}

\clearpage

\begin{figure}
\figurenum{2}
\includegraphics[angle=90,height=130mm]{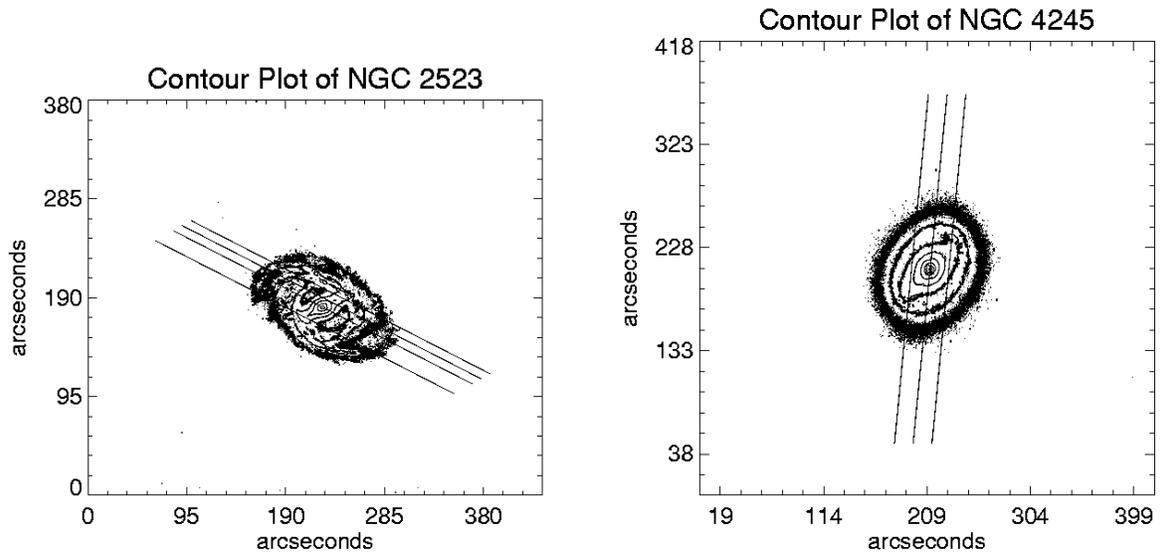}
\label{sample}
\caption{Contour plots derived from $B$-band images of NGC 2523 and NGC 4245 with the foreground stars removed. The images were taken from BCO. North is to the top and east is to the left. The lines indicate the slit positions used to obtain the stellar absorption-line spectra. For NGC 2523, the slits, from top to bottom, have a major axis offset of 17$\arcsec$ northwest, 9$\arcsec$ northwest, 0$\arcsec$, and 17$\arcsec$ southeast. For NGC 4245, the slits, from left to right, have a major axis offset of 17$\arcsec$ east, 0$\arcsec$, and 17$\arcsec$ west.}
\end{figure}

\clearpage

\begin{figure}
\figurenum{3}
\includegraphics[angle=0,height=125mm]{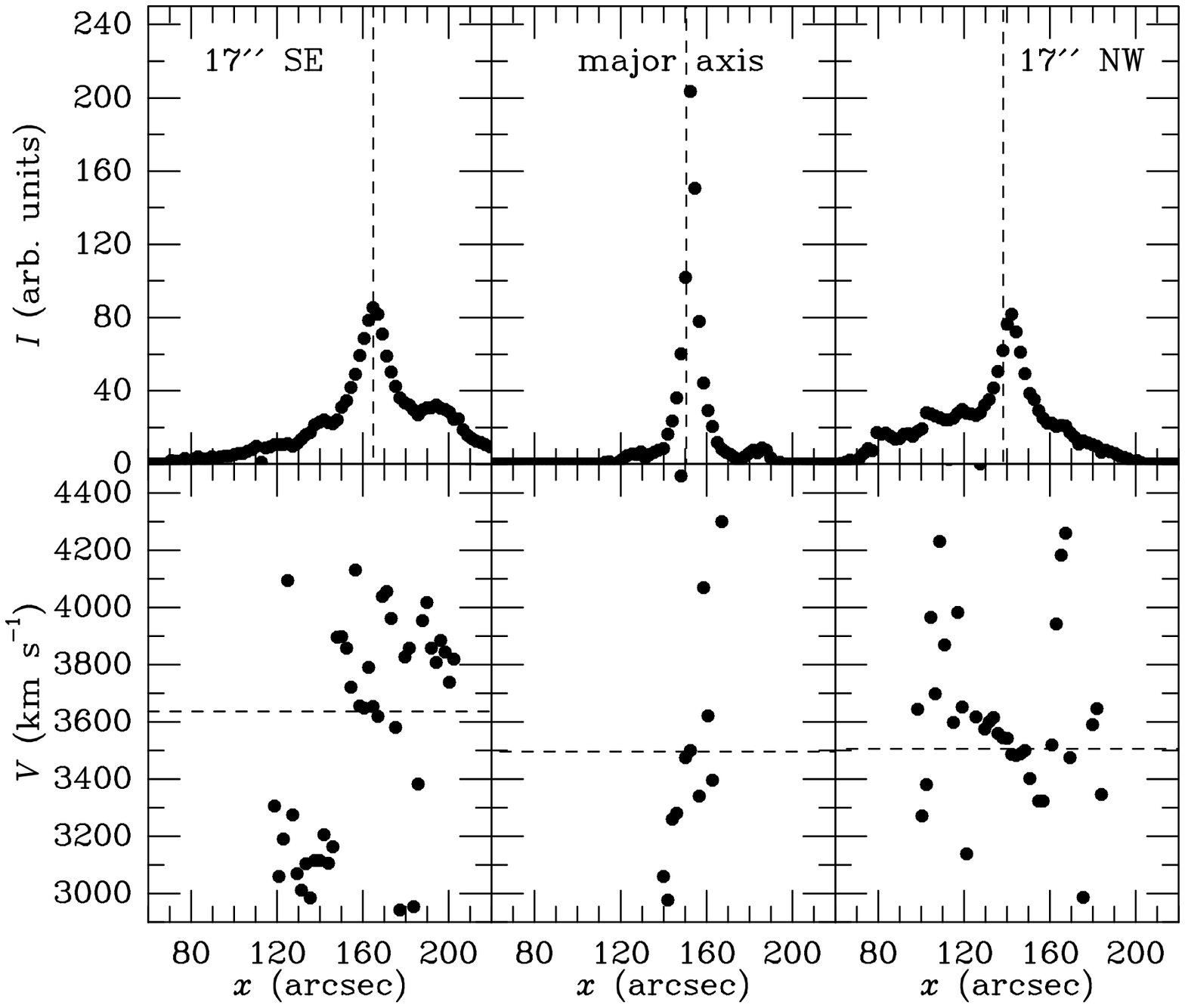}
\label{sample}
\caption{Luminosity ($I$) and velocity ($V$) profiles versus slit position $x$ for three slit positions of NGC 2523. The luminosity-weighted mean values for each profile are indicated by dashed lines. These means are also give in Table 1.}
\end{figure}

\clearpage

\begin{figure}
\figurenum{4}
\includegraphics[angle=0,height=125mm]{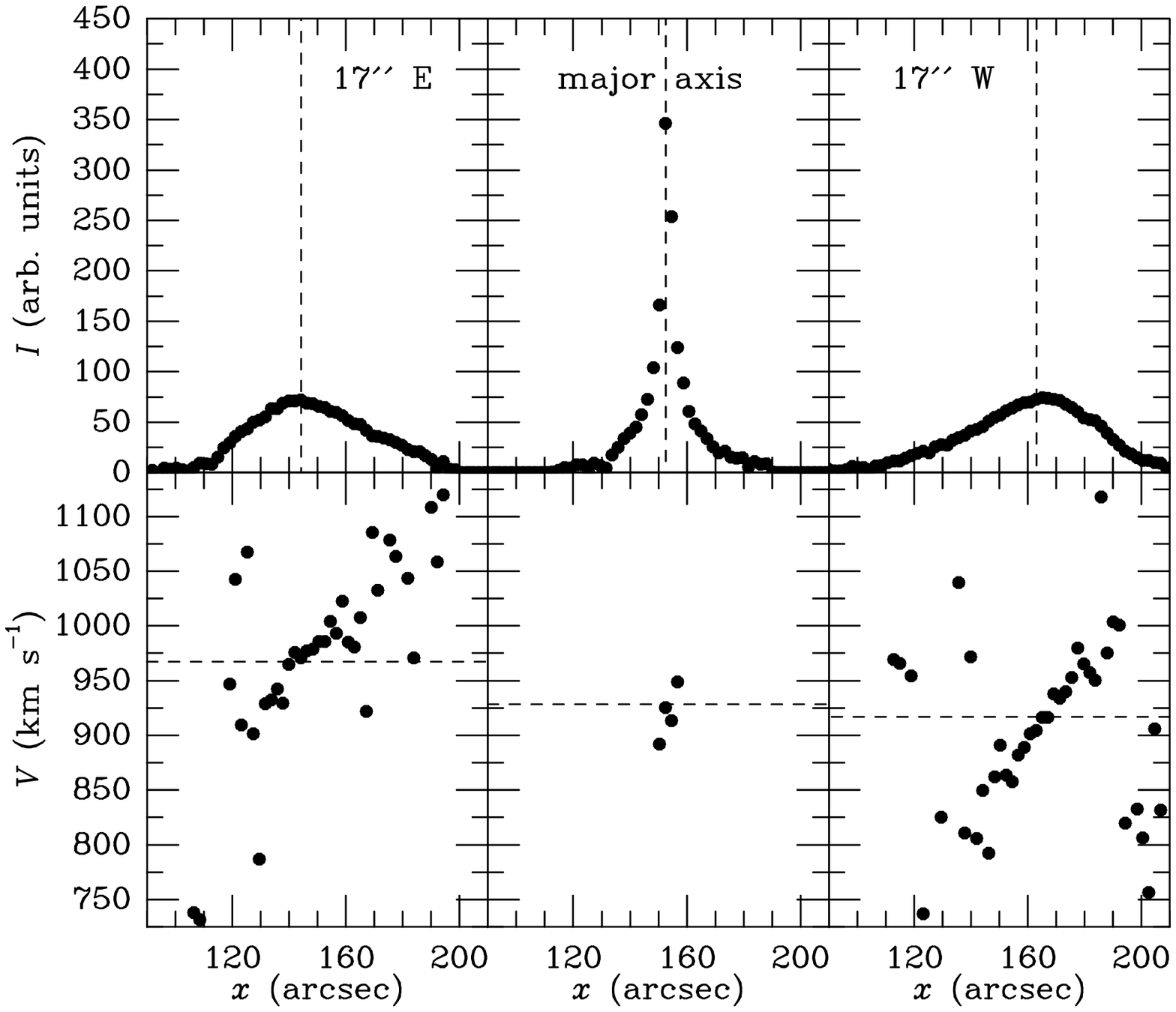}
\label{sample}
\caption{Luminosity ($I$) and velocity ($V$) profiles versus slit position $x$ for three slit positi
ons of NGC 4245. The luminosity-weighted mean values for each profile are indicated by dashed lines.
These means are also give in Table 1.}
\end{figure}

\clearpage

\begin{figure}
\figurenum{5}
\includegraphics[angle=90,height=130mm]{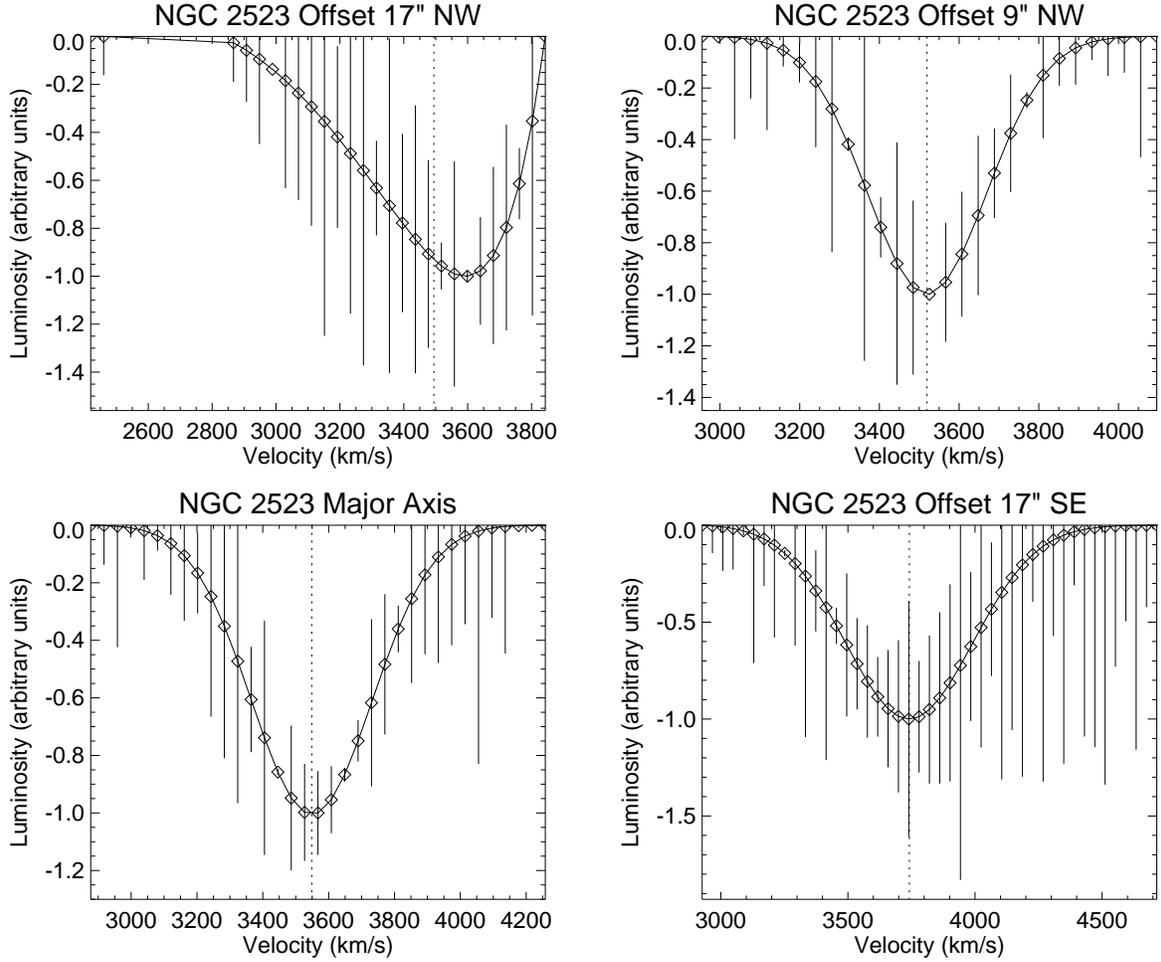}
\label{sample}
\caption{Examples of line profiles for the co-added 8542$\AA$ spectra for each offset position for NGC 2523. The error bars 
indicate the difference between the multiple Gaussian fit to the data and the actual data value. The dotted vertical line 
shows the mean of each distribution. The mean values for Offset 17$\arcsec$ NW, Offset 9$\arcsec$ NW, Major Axis, and 
Offset 17$\arcsec$ SE are 3494.2 $\pm$ 124.5, 3519.4 $\pm$ 83.9, 3548.8 $\pm$ 82.2, and 3741.8 $\pm$ 200.9 km s$^{-1}$, 
respectively.  }
\end{figure}

\clearpage

\begin{figure}
\figurenum{6}
\includegraphics[angle=90,height=130mm]{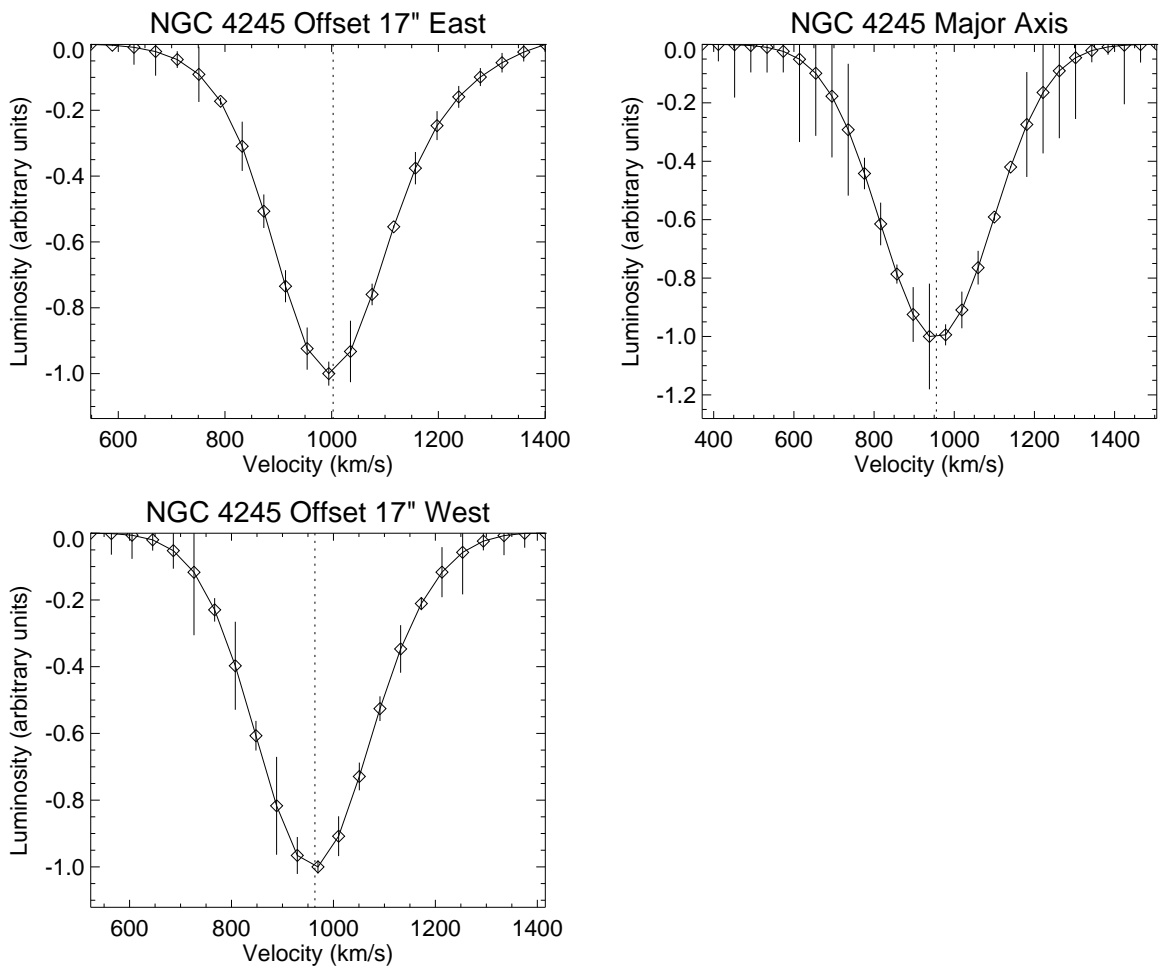}
\label{sample}
\caption{Examples of line profiles for the co-added 8542$\AA$ spectra for each offset position for NGC 4245. The error bars 
indicate the difference between the multiple Gaussian fit to the data and the actual data value. The dotted vertical line 
shows the mean of each distribution. The mean values for Offset 17$\arcsec$ E, Major Axis, and 
Offset 17$\arcsec$ W are 1002.7 $\pm$ 36.5, 955.4 $\pm$ 52.9, and 963.6 $\pm$ 41.9 km s$^{-1}$, 
respectively. }
\end{figure}

\clearpage

\begin{figure}
\figurenum{7}
\includegraphics[angle=90,height=125mm]{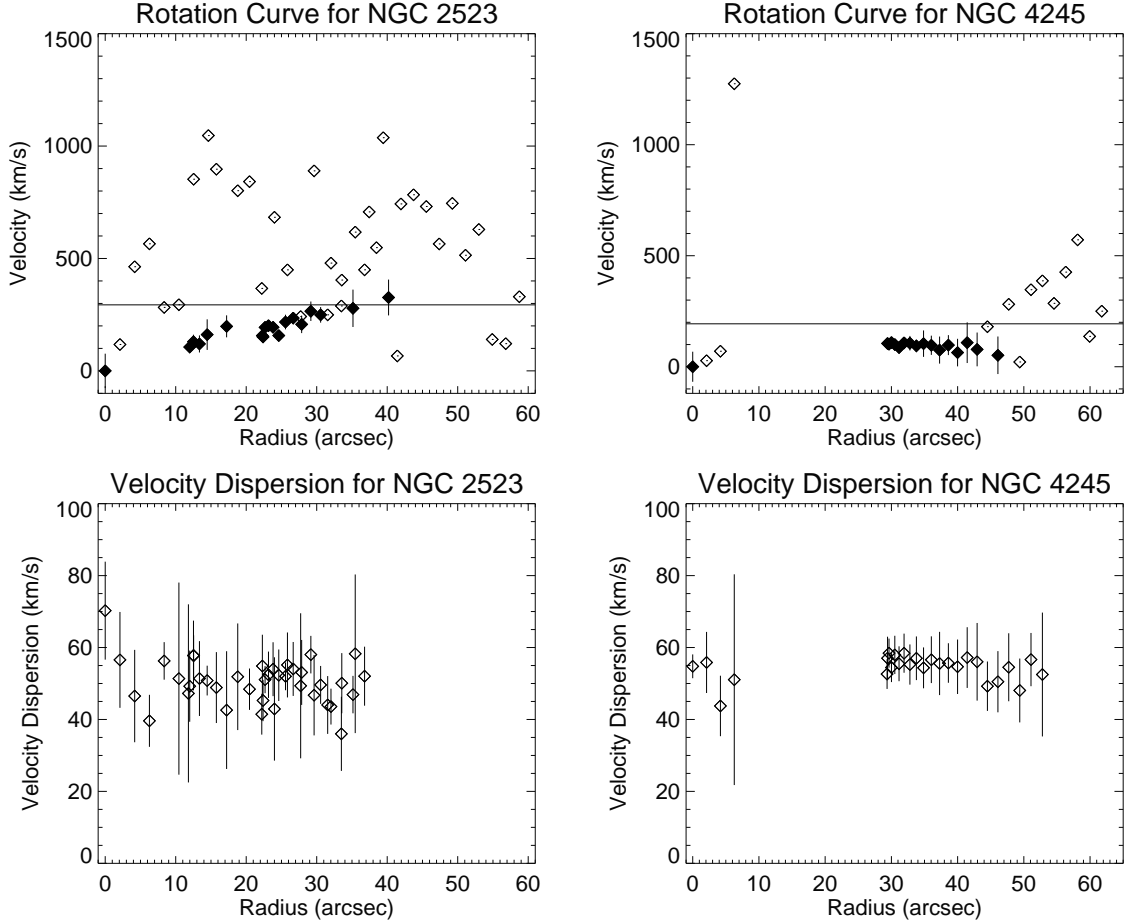}
\label{sample}
\caption{Rotation curves and velocity dispersions of NGC 2523 and NGC 4245 
derived from the velocities along each slit position. The filled rotation curve points 
represent data with errors less than 100 km s$^{-1}$. Error bars for rotation 
curve data points with errors greater than 100 km s$^{-1}$ are not shown for
the sake of clarity. XCOR was unable to determine rotational velocities beyond 
approximately 59$\arcsec$ for NGC 2523 and 62$\arcsec$
for NGC 4245. The horizontal line corresponds to the inclination corrected 
maximum gaseous circular velocity of 294 km s$^{-1}$ for NGC 2523 and 199 km s$^{-1}$ 
for NGC 4245. These maximum gaseous circular velocities were derived from the works of 
Kamphuis et al. (1996) and Garc\'{i}a-Barreto et al. (1994) respectively. The 
velocity dispersion values were corrected for an instrumental dispersion of 30
km s$^{-1}$. XCOR was unable to determine velocity dispersions beyond 
approximately 38$\arcsec$ for NGC 2523 and 55$\arcsec$ for NGC 4245.}
\end{figure}

\clearpage

\begin{figure}
\figurenum{8}
\includegraphics[angle=90,height=130mm]{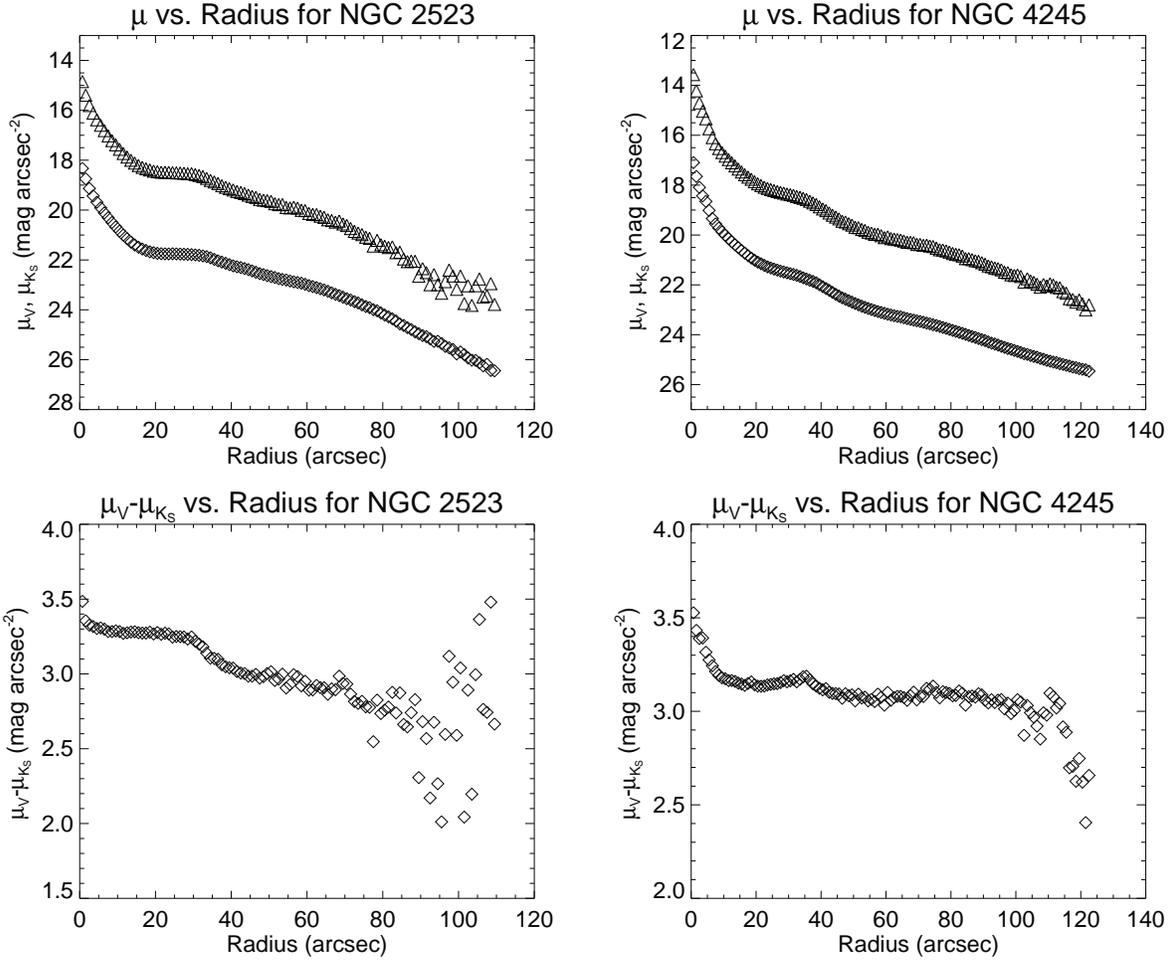}
\label{sample}
\caption{$V$ and $K_S$-band azimuthally
averaged surface brightness profiles of NGC 2523 and NGC 4245 (upper plots) and
the corresponding radial color profile for each galaxy (lower plots). The 
surface brightness profiles were derived using elliptical annulae with the
orientation parameters given in Table 2.
The diamonds in the surface brightness profiles represent $\mu_V$, while the 
triangles represent $\mu_{K_S}$.}
\end{figure}

\clearpage

\begin{figure}
\figurenum{9}
\includegraphics[angle=90,height=130mm]{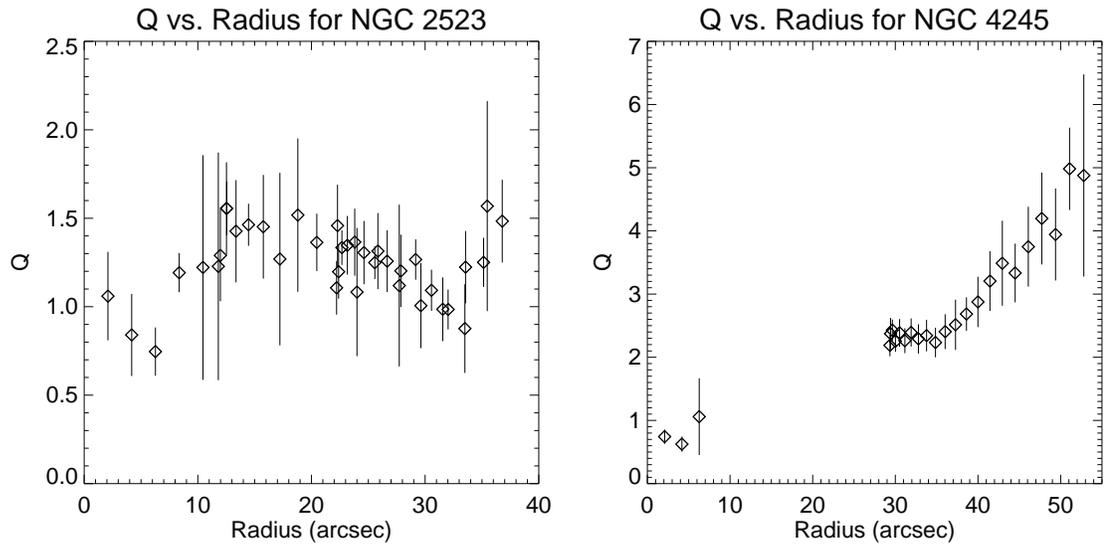}
\label{sample}
\caption{Plot of the Toomre stability parameter, $Q$, versus radius for 
NGC 2523 and NGC 4245 using the velocity 
dispersion data from this paper. For NGC 2523, $Q$ ranges from 0.7 $\pm$ 0.1 
to 1.6 $\pm$ 0.6. For NGC 4245, $Q$ ranges from 
0.6 $\pm$ 0.1 to 5.0 $\pm$ 0.7.}
\end{figure}

\clearpage

\begin{figure}
\figurenum{10}
\includegraphics[angle=90,height=130mm]{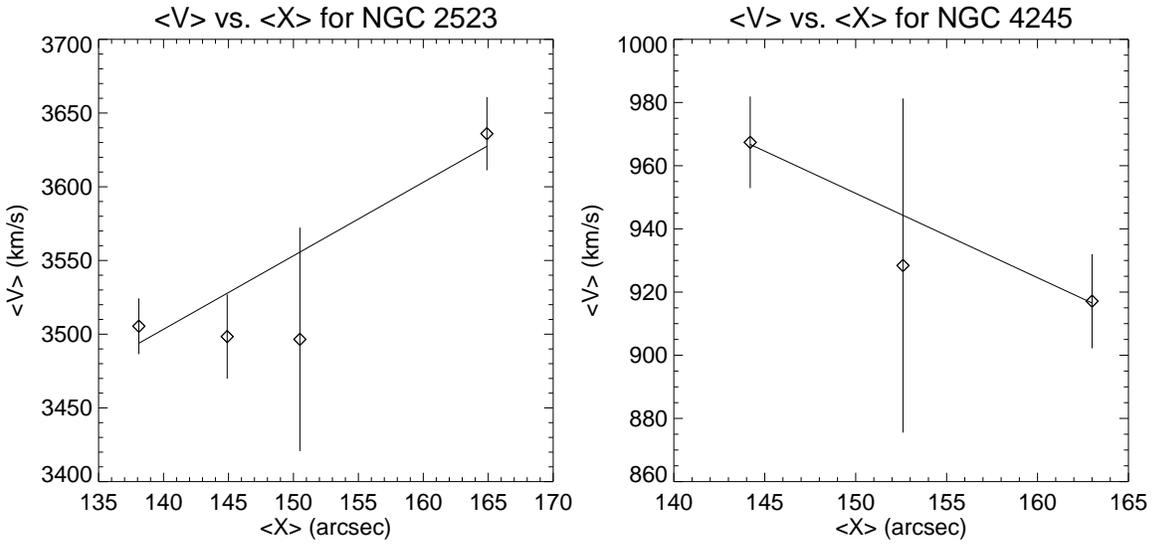}
\label{sample}
\caption{Plots of the luminosity weighted mean line-of-sight velocity as a function of the luminosity weighted mean position along the slit for both NGC 2523 and NGC 4245. The best-fitting regression lines weighted by the errors in $\langle$$V$$\rangle$ are also shown. The slope of the regression line gives the pattern speed as a function
of the galaxy's inclination.}
\end{figure}

\clearpage

\begin{figure}
\figurenum{11}
\includegraphics[angle=90,height=130mm]{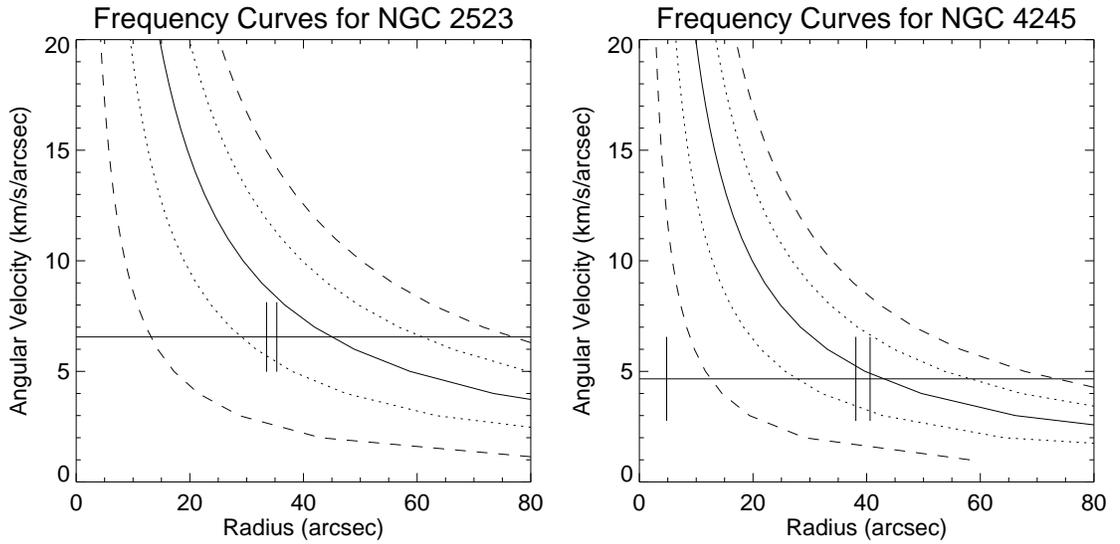}
\label{sample}
\caption{Frequency curves for NGC 2523 and NGC 4245. The horizontal lines correspond to the inclination corrected
$\Omega_{P}$ of 6.5 $\pm$ 1.5 km s$^{-1}$ arcsec$^{-1}$ for NGC 2523 and 4.6 $\pm$ 1.9 km s$^{-1}$ arcsec$^{-1}$ for NGC 4245. The short vertical
lines
correspond to the error range in $\Omega_{P}$. They are placed at the estimated bar (left) and inner ring (right) radii for NGC 2523 and the nuclear ring (left), bar (middle), and inner ring (right) radii for NGC 4245.
From left to right, the horizontal lines
intersect curves corresponding to $\Omega-\kappa$/2,
$\Omega-\kappa$/4, $\Omega$, $\Omega+\kappa$/4, and
$\Omega+\kappa$/2. $\Omega$ is the circular angular velocity and
$\kappa$ is the epicyclic frequency. The corotation radius is where
$\Omega$=$\Omega_{P}$.}
\end{figure}

\end{document}